\documentclass[11pt]{article}
\usepackage{graphicx}
\usepackage{cite}
\usepackage{epstopdf}
\usepackage{amssymb}
\usepackage{epstopdf}
\usepackage{hep}
\usepackage{feynmp}
\usepackage{multicol}
\usepackage{simplewick}
\usepackage{appendix}
\usepackage{subfigure}
\begin{document}

\title{{\bf Analysis of the double-spin asymmetry} {\boldmath $A_{LT}$} {\bf in \\
inelastic nucleon-nucleon collisions}}

\author{A.~Metz$^{1}$, D.~Pitonyak$^{1}$, A.~Sch\"afer$^{2}$, J.~Zhou$^{2}$
 \\[0.3cm]
{\normalsize\it $^1$Department of Physics, Barton Hall,
  Temple University, Philadelphia, PA 19122, USA} \\[0.15cm]
{\normalsize\it $^2$Institute for Theoretical Physics, Regensburg University,} \\ 
{\normalsize\it Universit\"atsstra{\ss}e 31, D-93053 Regensburg, Germany} \\[0.15cm]
}

\date{\today}
\maketitle

\begin{abstract}
\noindent
Within the collinear twist-3 framework, we analyze the double-spin asymmetry in collisions between longitudinally polarized nucleons and transversely polarized nucleons with focus on hadron and jet production.  As was the case in direct photon production, the double-spin dependent cross section for hadron and jet production has the advantage of involving a complete set of collinear twist-3 functions for a transversely polarized nucleon.  In addition, we outline further benefits of this observable for a potential future measurement at RHIC, which includes insight on the gluon helicity distribution as well as information on the Efremov-Teryaev-Qiu-Sterman function $T_{F}(x,x)$ that plays a crucial role in single-spin asymmetries.
\end{abstract}

%
%
%
\section{Introduction}
\label{s:intro}
Spin asymmetries in hard scattering processes have been an interesting subject of research for several decades.  Starting in the mid-1970s, the large single-spin asymmetries (SSAs) observed in inclusive hadron production \cite{Bunce:1976yb, Adams:1991rw, Krueger:1998hz, Adams:2003fx, Adler:2005in, :2008mi, Adamczyk:2012xd} were initially an obstacle for perturbative QCD.  Within the collinear parton model, these asymmetries should be on the order of  $\alpha_{s}m_{q}/P_{h\perp}$ \cite{Kane:1978nd, Ma:2008gm}, where $m_{q}$ is the mass of the quark, and $P_{h\perp}$ is the transverse momentum of the detected hadron.  However, research pioneered in the early 1980s  \cite{Efremov:1981sh} that went beyond the simplistic parton model showed that these SSAs could be generated within a framework that involved collinear twist-3 parton correlators.  This formalism, which is valid when a process contains one large scale $Q$ (with $\Lambda_{QCD}\ll Q$), has also been extensively investigated for SSAs in various observables --- see \cite{Qiu:1991pp, Qiu:1998ia, Eguchi:2006qz, Kouvaris:2006zy, Koike:2007rq, Zhou:2009jm, Koike:2009ge, Gamberg:2012iq, Metz:2012ui} for some specific examples.  (We also mention that other mechanisms have been proposed to explain large SSAs \cite{Hoyer:2006hu, Qian:2011ya, Kovchegov:2012ga}.)

Similarly, extensive work has been done on the longitudinal double-spin asymmetry (DSA) $A_{LL}$ in processes such as polarized lepton-nucleon collisions and polarized nucleon-nucleon collisions \cite{Bjorken:1969mm, Babcock:1978yc}.  This differs from the derivations of SSAs in that $A_{LL}$ is a leading twist (twist-2) effect that gives access to the helicity distributions of partons in the nucleon --- see \cite{deFlorian:2009vb} for a recent global extraction of these functions.  The main goal of this research has been to understand how the spin of the nucleon can be explained in terms of the partons that compose it.  A real surprise occurred when it was determined by EMC \cite{Ashman:1987hv} (and later confirmed at SLAC \cite{Anthony:1996mw, Abe:1998wq}) that the spins of the quarks contribute an unexpectedly small fraction to the spin of the nucleon.  Clearly, the remaining percentage must come from the orbital angular momentum of the partons and the spin of the gluons.  Much research has been done on this front to determine exactly what contribution each of these pieces make --- see, e.g., \cite{Kuhn:2008sy, Burkardt:2008jw, Aidala:2012mv} for recent reviews on the subject.

While the areas of hadronic spin physics outlined in the previous two paragraphs have for the most part operated independently of each other, one observable, namely, the longitudinal-transverse DSA $A_{LT}$, offers insight into both domains.  More specifically, $A_{LT}$ (in processes with one large scale) is a collinear twist-3 effect that is also sensitive to parton helicities.  The classic process that necessitates this formalism is $A_{LT}$ for inclusive deep-inelastic lepton-nucleon scattering (DIS).  In that case, one can study the collinear twist-3 function $g_{T}$.  In addition, $A_{LT}$ has been analyzed in the Drell-Yan process involving two incoming polarized hadrons \cite{Jaffe:1991kp, Tangerman:1994bb, Koike:2008du, Lu:2011th}.  More recently, $A_{LT}$ was calculated in inclusive lepton production from $W$-boson decay in proton-proton scattering \cite{Metz:2010xs}, for jet production in lepton-nucleon scattering \cite{Kang:2011jw}, and for direct photon production in nucleon-nucleon collisions \cite{Liang:2012rb}.  However, it was only in \cite{Liang:2012rb} that for the first time a spin dependent cross section was considered that required a complete set of collinear twist-3 functions for a transversely polarized nucleon in order to fully describe the observable.  We will see this same characteristic holds for hadron and jet production.  (Note that a term containing chiral-odd correlation functions was not computed in \cite{Liang:2012rb}.  We will also neglect these contributions in the present work --- see the discussion below Eq.~(\ref{e: collfac}).)  These higher-twist functions do not have a probability interpretation and are lesser known than the collinear ones relevant at leading twist (namely, the unpolarized distribution $f_{1}$, helicity distribution $g_{1}$, and transversity distribution $h_{1}$ \cite{Ralston:1979ys, Cortes:1991ja, Jaffe:1991kp}), but nevertheless they provide important insight into the spin structure of the nucleon. 

In the present work, we analyze the double-spin asymmetry $A_{LT}$ in nucleon-nucleon collisions for the case of hadron and jet production as well as review the results for direct photon production found in \cite{Liang:2012rb}.  These results collectively can be considered the DSA analog to the SSAs derived in the same processes \cite{Qiu:1991pp, Qiu:1998ia, Kouvaris:2006zy, Koike:2007rq}.  Furthermore, we briefly discuss plans for a future numerical study and highlight the prospects for this observable to provide insight on important areas of hadronic spin physics.  These include information on the gluon helicity distribution and the Efremov-Teryaev-Qiu-Sterman (ETQS) function $T_{F}(x,x)$ that enters into SSAs in hadronic processes.  

The paper is organized as follows:~in Sec.~\ref{s:T3}, we review the collinear twist-3 formalism including the relevant non-perturbative functions that enter into the calculation.  In Sec.~\ref{s:calccs}, we derive the double-spin dependent cross section for hadron and jet production, providing a few details of the calculation.  In addition, we briefly outline a future numerical study and emphasize potential benefits for a measurement of $A_{LT}$ at RHIC. In Sec.~\ref{s:sum}, we conclude the paper and summarize our work.

%
%
%
\section{Collinear twist-3 formalism}
\label{s:T3}
To start, let us make explicit the process under consideration, namely,
\begin{equation}
A(P,\,\vec{S}_{\perp}) + B(P',\,\Lambda) \rightarrow C(l) + X,
\end{equation}
where the 4-momenta and polarizations of the incoming nucleons $A$, $B$ and outgoing particle (or jet) $C$ are indicated.  The Mandelstam variables for the process are defined as $S = (P+P')^{2}$, $T = (P-l)^{2}$, and $U = (P'-l)^{2}$, which on the partonic level give $\hat{s} = xx' S$, $\hat{t} = xT/z$, and $\hat{u} = x' U/z$.  The longitudinal momentum fraction $x$ ($x'$) is associated with partons in the transversely (longitudinally) polarized nucleon.

The first non-vanishing contribution to the cross section is given by terms of twist-3 accuracy and reads 
\begin{align} \label{e: collfac}
d\sigma(\vec{l}_{\perp},\vec{S}_{\perp},\Lambda) &= \,H\otimes f_{a/A(3)}\otimes f_{b/B(2)}\otimes D_{C/c(2)} \nonumber \\
&+ \,H'\otimes f_{a/A(2)}\otimes f_{b/B(3)}\otimes D_{C/c(2)} \nonumber \\
&+ \,H''\otimes f_{a/A(2)}\otimes f_{b/B(2)}\otimes D_{C/c(3)},
\end{align} 
where a sum over partonic channels and parton flavors in each channel is understood.  In Eq.~(\ref{e: collfac}), $f_{a/A(t)}$ denotes the twist-$t$ distribution function associated with parton $a$ in hadron $A$ (and likewise for $f_{b/B(t)}$), while $D_{C/c(t)}$ represents the twist-$t$ fragmentation function associated with particle $C$ in parton $c$.  The factors $H$, $H'$, and $H''$ indicate the hard parts corresponding to each term, while the tensor product denotes convolutions in the appropriate momentum fractions.  For the case of the SSA $A_{UT}$ (where $B$ is now unpolarized), it has already been shown that the second term in (\ref{e: collfac}), which involves chiral-odd twist-3 unpolarized distributions, is negligible because of the smallness of the hard scattering coefficients \cite{Kanazawa:2000hz}.  We believe a similar statement will hold for the $A_{LT}$ case, which involves chiral-odd twist-3 helicity distributions, since the hard factors will be similar to the ones for the unpolarized case.  Arguments have been made that the first term in (\ref{e: collfac}) for $A_{UT}$ (so-called Sivers term) is dominant  \cite{Qiu:1991pp, Qiu:1998ia, Kouvaris:2006zy}.  However, recent work has shown that one cannot rule out significant contributions from the third term (so-called Collins term) \cite{Anselmino:2012rq, Kang:2010zzb, Kang:2011ni}.  For this current work on $A_{LT}$, we will focus on the first term in (\ref{e: collfac}) but cannot exclude that the third term could also play a critical role.  Therefore, for the situation we consider, $f_{b/B(2)} = g_{1}^{b}$ and $D_{C/c(2)} = D_{1}^{C/c}$, where $g_{1}$ and $D_{1}$ are the standard twist-2 helicity distribution function and unpolarized fragmentation function, respectively.  We then must determine what contributions are possible for $f_{a/A(3)}$.
\begin{figure}[t]
\begin{center}
\includegraphics[width=14cm]{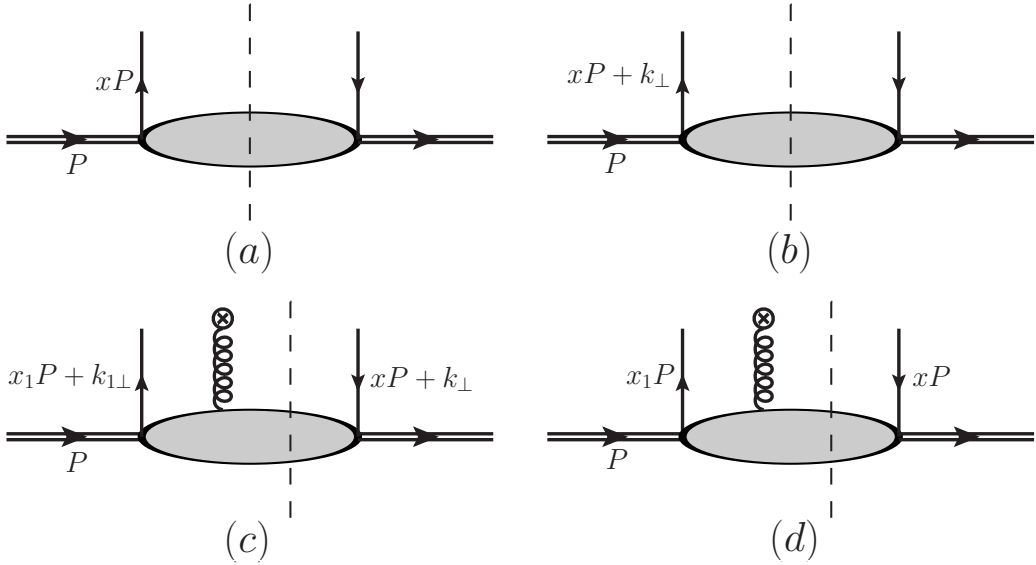}
\caption[]{Feynman diagrams for the twist-3 matrix elements that give contributions to $f_{a/A(3)}$.  See the text for more details.}
 \label{f:T3matrix}
\end{center}
\end{figure}

A detailed discussion of collinear twist-3 functions and, in particular, those relevant for a transversely polarized nucleon, is given in Ref.~\cite{Zhou:2009jm}.  Here we simply review the main aspects needed for this work.  The twist-3 matrix elements that we must consider are given by the diagrams in Fig.~\ref{f:T3matrix}.  Note that we have neglected matrix elements involving tri-gluon correlators.  In the lightcone ($A^{+} = 0$) gauge, these graphs lead to the three matrix elements \cite{Zhou:2009jm}
\begin{equation}
\langle\bar{\psi}\psi\rangle, \;\langle\bar{\psi}\partial_{\perp}\psi\rangle,\;\langle\bar{\psi}A_{\perp}\psi\rangle,
\end{equation}
which result from Figs.~\ref{f:T3matrix}(a), (b), and (d), respectively.  We do not have to consider Fig.~\ref{f:T3matrix}(c) because one does not need to simultaneously take into account $k_\perp$ expansion and $A_{\perp}$ gluon attachments (which would give rise to twist-4 contributions).  

Now that we have determined the relevant twist-3 matrix elements, we must parameterize them in terms of twist-3 functions that will eventually be involved in our final result.  We first focus on the quark-gluon-quark ($qgq$) matrix element $\langle\bar{\psi}A_{\perp}\psi\rangle$.  One notices that this matrix element is not gauge invariant.  This can be resolved in two ways:~rewrite the gluon field $A_{\perp}$ in terms of the field strength tensor $F^{+\mu}_{\perp} = \partial^{+}A_{\perp}^{\mu}$ or rewrite it in terms of the covariant derivative $D_{\perp}^{\mu} = \partial_{\perp}^{\mu}-igA_{\perp}^{\mu}$.  The former leads to the matrix element being written in terms of the so-called ``F-type'' functions, while the latter gives the so-called ``D-type'' functions \cite{Qiu:1991pp}.  Respectively, we have
\begin{eqnarray}
&& \int \frac{d\xi^-}{2 \pi} \, \frac{d\zeta^-}{2\pi} \, e^{ix_{1}P^+ \xi^-} e^{i(x-x_1)P^+\zeta^-}
\langle P,S_\perp | \bar{\psi}_\beta(0)  g F_\perp^{+ \mu}(\zeta^-)\psi_\alpha(\xi^{-}) | P,S_\perp \rangle
\nonumber \\
&& \hspace{0.5cm}
= \frac{M}{2} \Big[ F_{FT}(x,x_1) \, \epsilon^{\mu\nu}_\perp S_{\perp \nu} \slashed{n}
- G_{FT}(x,x_1) \, i S_\perp^\mu \gamma_5 \slashed{n} \Big]_{\alpha\beta}, 
\label{e:F-type}
\end{eqnarray}
and
\begin{eqnarray}
&& \int \frac{d\xi^-}{2 \pi} \frac{d\zeta^-}{2\pi} \, e^{ix_{1}P^+ \xi^-} \, e^{i(x-x_1)P^+\zeta^-}
\langle P,S_\perp | \bar{\psi}_\beta(0)  i D_\perp^\mu(\zeta^-)\psi_\alpha(\xi^{-}) | P,S_\perp \rangle
\nonumber \\
&& \hspace{0.5cm}
= \frac{M}{2P^+} \Big[ F_{DT}(x,x_{1}) \, i\epsilon^{\mu\nu}_\perp S_{\perp\nu} \slashed{n}
+G_{DT}(x,x_{1}) \, S_\perp^\mu \gamma_5 \slashed{n} \Big]_{\alpha\beta}.
\label{e:D-type}
\end{eqnarray}
In Eqs.~(\ref{e:F-type}), (\ref{e:D-type}), we have suppressed Wilson lines and have indicated the nucleon mass by $M$.  We have also introduced the lightcone vector $n = (1^+,0^-,\vec{0}_\perp)$, whose conjugate vector is $\bar{n} = (0^+,1^-,\vec{0}_\perp)$.  Note that we have defined the F-type and D-type functions as in Ref.~\cite{Meissner:2006}, which differs from those used in \cite{Liang:2012rb}. These functions satisfy certain symmetry properties under the interchange of their arguments:
\begin{equation} \label{e:Fsym}
F_{FT}(x,\,x_{1})=F_{FT}(x_{1},\,x)\;\;{\rm and}\;\; G_{FT}(x,\,x_{1})=-G_{FT}(x_{1},\,x),
\end{equation}
while
\begin{equation} \label{e:Dsym}
F_{DT}(x,\,x_{1})=-F_{DT}(x_{1},\,x)\;\; {\rm and}\;\; G_{DT}(x,\,x_{1})=G_{DT}(x_{1},\,x).
\end{equation}

Moreover, it turns out the F-type and D-type functions are not independent of each other.  One can establish the following relations between these functions \cite{Eguchi:2006qz}:
\begin{eqnarray}
&& F_{DT}(x,x_{1}) = PV \frac{1}{x-x_{1}} \, F_{FT}(x,x_{1}),
\label{e:DF_1}
\\
&& G_{DT}(x,x_{1})= PV \frac{1}{x-x_{1}} \, G_{FT}(x,x_{1}) + \delta(x-x_{1}) \, \tilde{g}(x),
\label{e:DF_2}
\end{eqnarray}
where $PV$ denotes the principal value.  In order to derive these expressions, notice that we must introduce an additional twist-3 function $\tilde{g}(x)$, whose definition is given by
\begin{eqnarray}
&& \int \frac{d\xi^-}{2 \pi} \, e^{ixP^+ \xi^-}
\langle P,S_\perp | \bar{\psi}_\beta(0)  
\bigg( iD_\perp^\mu(\xi^{-}) + g\int_{\xi^{-}}^\infty d\zeta^- F_\perp^{+\mu}(\zeta^-) \bigg)
\psi_\alpha(\xi^{-}) | P,S_\perp \rangle
\nonumber\\
&& \hspace{0.5cm}
= \frac{M}{2} \Big[ \tilde{g}(x) \, S_\perp^\mu \gamma_5 \slashed{n} \Big]_{\alpha\beta}. 
\end{eqnarray}
This function is associated with the quark-quark ($qq$) matrix element $\langle\bar{\psi}\partial_{\perp}\psi\rangle$.  We also mention that $\tilde{g}(x)$ is equivalent to the first $k_{\perp}$-moment of the TMD $g_{1T}(x,\vec{k}_{\perp}^{2})$ for a longitudinally polarized quark in a transversely polarized nucleon \cite{Zhou:2009jm}:
\begin{equation}
\tilde{g}(x) = \int d^2 \vec{k}_{\perp} \, \frac{\vec{k}_{\perp}^{\,2}}{2 M^2} \, 
g_{1T}(x,\vec{k}_{\perp}^{\,2}).
 \label{e:g1T}
\end{equation}
The other relevant $qq$ matrix element $\langle\bar{\psi}\psi\rangle$ leads to a contribution from the well-known twist-3 function $g_{T}(x)$, whose definition is given by
\begin{equation} \label{e: g_T}
\frac{2M} {P^{+}}\,S_{\perp}^{\mu}\,g_{T}(x)= \int \frac{d y^{-}} {2\pi}\, e^{i xP^{+} y^{-}}\, \langle P,\,S_{\perp}| \bar{\psi}(0)\gamma^{\mu}\gamma^{5}\psi(y^{-})|P,\,S_{\perp}\rangle.
\end{equation}
However, $g_{T}(x)$ can be related to the D-type functions (and, therefore, due to (\ref{e:DF_1}), (\ref{e:DF_2}), also the F-type functions) through the QCD equations of motion (EOM) \cite{Efremov:1981sh, Jaffe:1991kp}:
\begin{equation} \label{e:EOMgT}
x\,g_{T}(x)=\int dx_{1}\, [G_{DT}(x,\,x_{1})-F_{DT}(x,\,x_{1})].
\end{equation}

From the above discussion, we have identified six twist-3 functions relevant for a transversely polarized nucleon: $\tilde{g},\,g_{T},\,F_{FT},\,G_{FT},\,F_{DT},\,G_{DT}$.  However, from the relations given in Eqs.~(\ref{e:DF_1}), (\ref{e:DF_2}), (\ref{e:EOMgT}), in the end one has only three independent collinear twist-3 functions relevant for a transversely polarized nucleon.  At the outset of a calculation, one can choose to work with either the F-type functions and $\tilde{g}(x)$ or the D-type functions and $\tilde{g}(x)$.  One cannot simply use the F-type or D-type functions alone, but rather the function $\tilde{g}(x)$ must also be included --- see, e.g., Ref.~\cite{Boer:1997bw}.
%
%
%
\section{Calculation of the double-spin dependent cross section}
\label{s:calccs}

\subsection{General structure of the calculation} 
\label{s:general}

\begin{figure}[t] 
\begin{center}
\includegraphics[width=17cm]{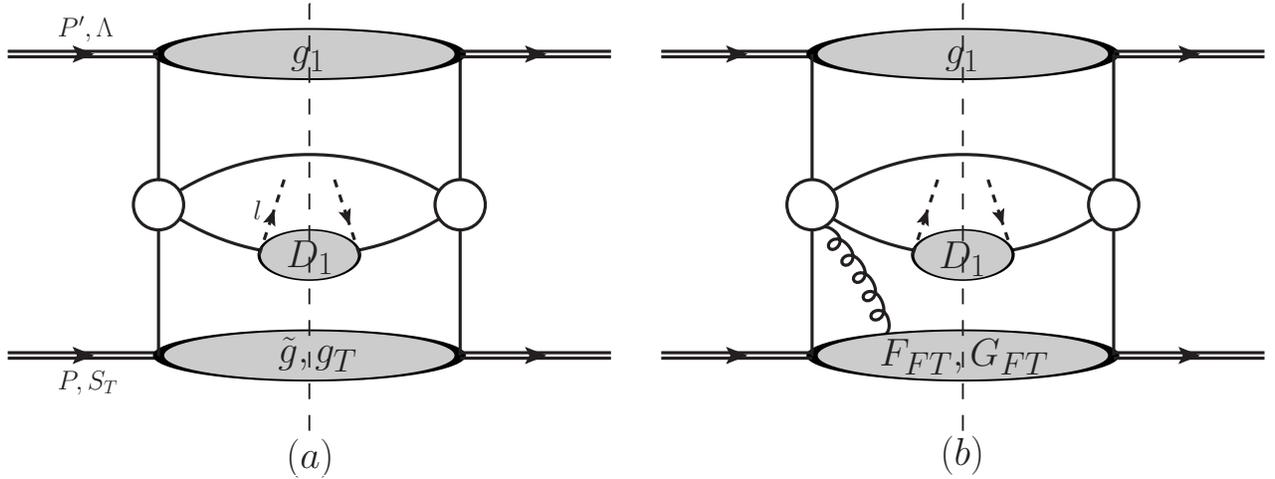}
\caption[] {Graphs showing factorization for contributions to $A_{LT}$ from (a) $qq$ correlators and (b) $qgq$ correlators.} \label{f: A_LT_fact}
\end{center}
\end{figure}

The factorization of the process under consideration is shown in Fig.~\ref{f: A_LT_fact}.  This includes collinear factors associated with the longitudinally polarized nucleon (top gray blob), the outgoing particle (or jet) (middle gray blob), and the transversely polarized nucleon (bottom gray blob) as well as hard factors (white blobs).  We choose to work with the F-type functions and $\tilde{g}(x)$.  For each partonic channel, the main task becomes calculating the hard scattering coefficients for each of these functions, which then allows us to write down the double-spin dependent cross section.  We will denote each channel by $ab\rightarrow cd$, where $a$$\,$($b$) is the parton associated with the transversely (longitudinally) polarized nucleon and $c$ is the parton that fragments into the detected particle (or jet).   

\begin{figure}[t]  
\begin{center}
\includegraphics[width=12cm]{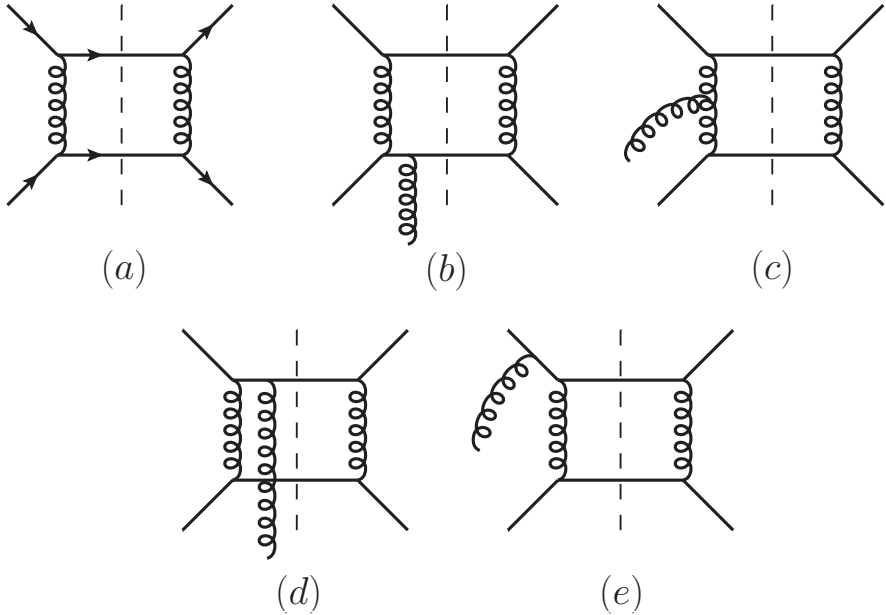}
\caption[]{Hard scattering diagrams for the $qq' \rightarrow qq'$ channel involving (a) $qq$ correlators and (b)--(e) $qgq$ correlators.  Note that Hermitian conjugate diagrams for the $qgq$ graphs are not shown.} 
\label{f:qqprime_qqprime}
\end{center}
\end{figure}

Here we will focus on the $qq'\rightarrow qq'$ channel in order to present a few details of the calculation.  The relevant hard scattering diagrams for this channel are shown in Fig.~\ref{f:qqprime_qqprime}.  First, we consider the graph in Fig.~\ref{f:qqprime_qqprime}(a).  If we keep the transverse momentum of the initial state parton $q$ (as in Fig.~\ref{f:T3matrix}(b)), then we can determine the hard scattering coefficient for $\tilde{g}(x)$.  In fact, this will lead to terms involving both $\tilde{g}(x)$ and its derivative, as was first detailed in \cite{Qiu:1991pp, Qiu:1998ia}.  On the other hand, if we neglect the transverse momentum of $q$ in the initial state (as in Fig.~\ref{f:T3matrix}(a)), then we obtain the hard part for $g_{T}(x)$.  However, since we work with the F-type functions and $\tilde{g}(x)$, we use Eq.~(\ref{e:EOMgT}) in conjunction with Eqs.~(\ref{e:DF_1}), (\ref{e:DF_2}) to write $g_{T}(x)$ in terms of those functions.  Lastly, we must attach gluons in all possible ways to Fig.~\ref{f:qqprime_qqprime}(a), which leads to Figs.~\ref{f:qqprime_qqprime}(b)--(e) and their Hermitian conjugates (not shown).  In these graphs we can neglect the transverse momenta of the initial state parton $q$ and gluon (as in Fig.~\ref{f:T3matrix}(d)).  These diagrams allow us to find the $qgq$ contributions to the hard factors for the F-type functions.  Note that we can combine the graphs in Figs.~\ref{f:qqprime_qqprime}(b)--(e) with their Hermitian conjugates by using the symmetry relations in Eq.~(\ref{e:Fsym}).  

We remark at this stage that in general the $qgq$ diagrams are not always real but can acquire an imaginary part whenever internal parton lines go on-shell.  This requires the use of the distribution identity 
\begin{equation}
\frac{1}{x \pm i\epsilon}=PV \frac{1}{x} \mp i \pi \delta(x).
\end{equation}
However, unlike the case of SSAs, the $PV$ part survives when we combine the various cut diagrams, whereas the pole term vanishes --- see also \cite{Liang:2012rb}.  A related feature is that \emph{all} of the $qgq$ graphs contribute to the hard scattering coefficients for the F-type functions, unlike the situation for SSAs when one considers the so-called soft gluon pole (SGP) term.  For example, if one were calculating the SGP term to the SSA for $AB\rightarrow CX$ for the $qq'\rightarrow qq'$ channel, only Figs.~\ref{f:qqprime_qqprime}(b), (e) (after including the Hermitian conjugate graphs) provide such a pole \cite{Kouvaris:2006zy}.

Finally, collecting all the terms, we find for the $qq'\rightarrow qq'$ channel the following contribution to the double-spin dependent cross section:
\begin{eqnarray}
\frac{l^{0}d\sigma^{qq'\rightarrow qq'}(\vec{S}_{\perp},\,\Lambda)} {d^{3}\vec{l}} \!\!\!&=&\!\!\! -\frac{2\alpha_{s}^{2}M} {S} \vec{l}_{\perp}\cdot\vec{S}_{\perp}\,\Lambda\sum_{a,\,b,\,c}\,\int_{z_{min}}^{1}\frac{dz} {z^{2}}\,D_{1}^{C/c}(z)\,\int_{x'_{min}}^{1}\,\frac{dx'} {x'}\frac{1} {x' S+T/z}\frac{1} {z\,\hat{u}}\,g_{1}^{b}(x')\,\frac{1} {x} \nonumber \\[0.3cm]
&&\hspace{-0.6cm}\times\,\left\{\tilde{g}^{a}(x)\left[\frac{C_{F}} {N_{c}}\frac{2\hat{s}^{2}-\hat{t}^{2}} {\hat{t}^{2}}\right]-x\frac{d\tilde{g}^{a}(x)} {dx}\left[\frac{C_{F}} {N_{c}}\frac{\hat{u}-\hat{s}} {\hat{t}}\right]\right. \nonumber \\[0.3cm]
&&\hspace{-0.6cm}+\,\int dx_{1}\,PV\frac{1} {x-x_{1}}G_{FT}^{a}(x,\,x_{1})\,\left[\frac{1} {2}\left(\frac{2(\hat{u}-\hat{s})} {\xi\hat{t}} +\frac{\hat{s}(\hat{s}-\hat{u})} {(1-\xi)\hat{t}^{2}}+
\frac{\hat{u}} {\hat{t}}\right)\right.\nonumber\\[0.3cm]
&&\hspace{4.92cm}+\,\left.\frac{1} {2N_{c}^{2}}\left(\frac{2(\hat{s}-\hat{u})} {\hat{t}}\left(\frac{2\hat{t}-\hat{u}} {\xi\hat{t}}+\frac{1} {1-\xi}\right)-\frac{\hat{u}} {\hat{t}}\right)\right]\nonumber \\[0.3cm]
&&\hspace{-0.6cm}-\left.\,\int dx_{1}\,PV\frac{1} {x-x_{1}}F_{FT}^{a}(x,\,x_{1})\left[\frac{1} {2} \left(\frac{\hat{u}} {\hat{t}}-\frac{\hat{s}(\hat{s}-\hat{u})} {(1-\xi)\hat{t}^{2}}\right) + \frac{1} {2N_{c}^{2}}\left(-\frac{2(\hat{s}-\hat{u})} {\hat{t}(1-\xi)}-\frac{\hat{u}} {\hat{t}}\right)\right]\right\},\nonumber\\ \label{e:sigqqprimeF}
\end{eqnarray}
where $x=-x'(U/z)/(x'S+T/z)$, $x'_{min}=-(T/z)/(U/z+S)$, and $z_{min}=-(T+U)/S$.  We have introduced $\xi = x_{g}/x$, where $x_{g}=x-x_1$, and understand $1/\xi$ to mean $PV(1/\xi)$.  The $SU(3)$ color factors depend on $C_{F} =4/3$ and $N_{c}=3$.  We note that the coefficient of $(d/dx)\tilde{g}(x)$ in Eq.~(\ref{e:sigqqprimeF}) matches the hard factor for the $qq'\rightarrow qq'$ channel in the leading order (LO) calculation of $A_{LL}$ found in \cite{Babcock:1978yc}.  This is to be expected given the Dirac projectors associated with $g_{1}^{b}(x)$ and $\tilde{g}^{a}(x)$ and the fact the ``derivative term'' at this stage is obtained by neglecting transverse momentum everywhere except in the on-shell delta function \cite{Qiu:1991pp, Qiu:1998ia}.  We have checked for all channels that at this point in the calculation an agreement occurs between the derivative term and the LO $A_{LL}$ coefficients.

We can rewrite (\ref{e:sigqqprimeF}) in terms of the D-type functions and $\tilde{g}(x)$ by using Eqs.~(\ref{e:DF_1}), (\ref{e:DF_2}).  If one does so, a nice simplification occurs involving $\tilde{g}(x)$ and its derivative: 
\begin{eqnarray}
\frac{l^{0}d\sigma^{qq'\rightarrow qq'}(\vec{S}_{\perp},\,\Lambda)} {d^{3}\vec{l}} \!\!\!&=&\!\!\! -\frac{2\alpha_{s}^{2}M} {S} \vec{l}_{\perp}\cdot\vec{S}_{\perp}\,\Lambda\sum_{a,\,b,\,c}\,\int_{z_{min}}^{1}\frac{dz} {z^{2}}\,D_{1}^{C/c}(z)\,\int_{x'_{min}}^{1}\,\frac{dx'} {x'}\frac{1} {x' S+T/z}\frac{1} {z\,\hat{u}}\,g_{1}^{b}(x')\,\frac{1} {x} \nonumber \\[0.3cm]
&&\times\,\left\{\left(\tilde{g}^{a}(x)-x\frac{d\tilde{g}^{a}(x)} {dx}\right)\left[-\frac{1} {2N_{c}^{2}}\frac{(\hat{t}-\hat{u})(\hat{s}-\hat{u})} {\hat{t}^{2}}\right]\right. \nonumber \\[0.3cm]
&&+\,\int dx_{1}\,G_{DT}^{a}(x,\,x_{1})\,\left[\frac{1} {2}\left(\frac{2(\hat{u}-\hat{s})} {\xi\hat{t}} +\frac{\hat{s}(\hat{s}-\hat{u})} {(1-\xi)\hat{t}^{2}}+
\frac{\hat{u}} {\hat{t}}\right)\right.\nonumber\\[0.3cm]
&&\hspace{3.73cm}+\,\left.\frac{1} {2N_{c}^{2}}\left(\frac{2(\hat{s}-\hat{u})} {\hat{t}}\left(\frac{2\hat{t}-\hat{u}} {\xi\hat{t}}+\frac{1} {1-\xi}\right)-\frac{\hat{u}} {\hat{t}}\right)\right]\nonumber \\[0.3cm]
&&-\left.\,\int dx_{1}F_{DT}^{a}(x,\,x_{1})\left[\frac{1} {2} \left(\frac{\hat{u}} {\hat{t}}-\frac{\hat{s}(\hat{s}-\hat{u})} {(1-\xi)\hat{t}^{2}}\right) + \frac{1} {2N_{c}^{2}}\left(-\frac{2(\hat{s}-\hat{u})} {\hat{t}(1-\xi)}-\frac{\hat{u}} {\hat{t}}\right)\right]\right\}.\nonumber\\ \label{e:sigqqprimeD}
\end{eqnarray}
We will comment more on this ``compact'' form involving $\tilde{g}(x)$ and its derivative as well as make other general remarks on the analytical result in the next subsection.

\subsection{Final analytical result}
\label{s:final}
Following for the remaining channels the outline given above for calculating hard factors, we find the cross section relevant for the DSA $A_{LT}$ in $AB\rightarrow C X$ is given by
\begin{eqnarray}
\frac{l^{0}d\sigma(\vec{S}_{\perp},\,\Lambda)} {d^{3}\vec{l}} \!\!\!&=&\!\!\! -\frac{2\alpha_{s}^{2}M} {S} \vec{l}_{\perp}\cdot\vec{S}_{\perp}\,\Lambda\sum_{i}\sum_{a,\,b,\,c}\,\int_{z_{min}}^{1}\frac{dz} {z^{2}}\,D_{1}^{C/c}(z)\,\int_{x'_{min}}^{1}\,\frac{dx'} {x'}\frac{1} {x' S+T/z}\frac{1} {z\,\hat{m}_{i}}\,g_{1}^{b}(x')\,\frac{1} {x} \nonumber \\[0.3cm]
&&\times\,\left\{\left[\tilde{g}^{a}(x)-x\frac{d\tilde{g}^{a}(x)} {dx}\right]\,H_{\tilde{g}}^{i}+\int dx_{1}\left[G_{DT}^{a}(x,\,x_{1})\,H_{G_{DT}}^{i}-F_{DT}^{a}(x,\,x_{1})\,H_{F_{DT}}^{i}\right]\right\},\nonumber\\ \label{e:sigmahadron}
\end{eqnarray}
where $i$ denotes the channel and $\hat{m}_{i}$ the corresponding partonic Mandelstam variable for that channel (see Table \ref{t:1} in Appendix \ref{a:hadron}).  The result in Eq.~(\ref{e:sigmahadron}) is if the detected particle is a hadron, with the hard scattering coefficients $H^{i}$ given in Appendix \ref{a:hadron}.  However, one can also obtain the expression for the double-spin dependent cross section for jet production by setting $D_{1}^{C/c}(z) = \delta(1-z)$.  The hard parts in this case are again given in Appendix \ref{a:hadron}, but now one can combine channels that differ by a crossing of the final state partons.  Likewise, for direct photon production one must set $D_{1}^{c}(z) = \delta(1-z)$ but also must make the replacement $\alpha_{s}\rightarrow \alpha_{em} e_{a}^{2}$ for one factor of $\alpha_{s}$, where $e_{a}$ is the charge of a quark with flavor $a$ (in units of $e$).  In this case, the number of channels reduces significantly, and their respective hard parts, which first appeared in \cite{Liang:2012rb}, are given in Appendix \ref{a:photon}.  We note a correction in the overall sign for the hard factors for the $qg\rightarrow \gamma q$ channel.

A few comments are in order on the analytical result.  First, we again mention that this calculation of $A_{LT}$ is the analog to the calculation of $A_{UT}$ in the same processes\cite{Qiu:1991pp, Qiu:1998ia, Kouvaris:2006zy, Koike:2007rq}.  Second, as we saw with the $qq'\rightarrow qq'$ channel, when we write the result using the D-type functions instead of the F-type functions, $\tilde{g}(x)$ and its derivative combine in the same compact form as $T_{F}(x,x)$ did for SSAs in direct photon and inclusive pion production \cite{Qiu:1991pp, Kouvaris:2006zy, Koike:2007rq}.  This form was also seen in the $A_{LT}$ calculations done in Refs.~\cite{Metz:2010xs, Kang:2011jw} (and, as mentioned above, Ref.~\cite{Liang:2012rb}).  Finally, we see Eq.~(\ref{e:sigmahadron}) involves a complete set of collinear twist-3 functions for a transversely polarized nucleon.  This is because the hard parts associated with each contribution are not the same, and, therefore, we cannot combine them into a simpler function.  Thus, in principle this process in conjunction with other reactions allows us to access a complete set of collinear twist-3 functions for a transversely polarized nucleon.

At this point, we would also like to make a few observations about the hard scattering coefficients $H^{i}_{F_{DT}}$ and $H^{i}_{G_{DT}}$ in Appendices \ref{a:hadron} and \ref{a:photon}.  We remark that these hard parts, as we have made explicit in the Appendices, can all be written in the form
\begin{equation}
H^{i} = H^{i}_{1} + \frac{1} {1-\xi} \,H^{i}_{2} + \frac{1} {\xi}\, H^{i}_{3},
\label{e:Hterms}
\end{equation}
where $H_{1}^{i}$, $H_{2}^{i}$, and $H_{3}^{i}$ are independent of $\xi$, and we have dropped the $F_{DT}$ and $G_{DT}$ subscripts from the $H$'s.  First, notice that $H_{1,\,F_{DT}}^{i}$ = $H_{1,\,G_{DT}}^{i}$, which means one could pull these hard factors out of the integral over $x_{1}$ and, using (\ref {e:EOMgT}), write a term involving $g_{T}(x)$.  For some channels, like $qq'\rightarrow qq'$, this is a trivial statement because the only $\xi$-independent terms in $H^{i}_{F_{DT}}$ and $H^{i}_{G_{DT}}$ come from the $g_{T}(x)$ contribution --- see the second paragraph of Sec.~\ref{s:general}.  However, for other channels, like $qg\rightarrow qg$, the $qgq$ contributions to $H^{i}_{F_{DT}}$ and $H^{i}_{G_{DT}}$ also contain $\xi$-independent terms.  We find it interesting that these additional $\xi$-independent terms are always the same for $H^{i}_{F_{DT}}$ and $H^{i}_{G_{DT}}$.  Second, one sees that $H^{i}_{2,\,F_{DT}}$ = $-H^{i}_{2,\,G_{DT}}$ and $H^{i}_{3,\,F_{DT}} = 0$.  We leave the former as another intriguing observation on the structure of the result.  For the latter, we remark that the pole contribution $1/\xi$ comes from initial/final state interactions and can be written as a kinematical factor times the Born cross section --- see, e.g., \cite{Qiu:1991pp, Qiu:1998ia, Kouvaris:2006zy}.  In this case, the Born diagram corresponding to $F_{DT}$ vanishes, which leads to $H^{i}_{3,\,F_{DT}} = 0$.  Finally, one notices that $H^{ab\rightarrow cd}_{F_{DT}} = -H^{ab\rightarrow dc}_{F_{DT}}(\hat{t} \leftrightarrow \hat{u})$, and similarly for $H^{ab\rightarrow cd(dc)}_{G_{DT}}$, where $(\hat{t}\leftrightarrow\hat{u})$ means interchange of $\hat{t}$ and $\hat{u}$.  Since one might surmise these hard parts for $ab\rightarrow dc$ can be obtained from those for $ab\rightarrow cd$ by an interchange of $\hat{t}$ and $\hat{u}$ (because we neglect $k_{\perp}$ and $k_{1\perp}$), the negative sign might be a bit unexpected.  This negative sign appears to be due to the sensitivity of the transversely polarized gluon attachments to the transverse momentum of the outgoing partons.  When one interchanges final state partons, the transverse momenta of the outgoing partons change signs, which is reflected in the crossed-channel hard parts.  Also, the fact that $H^{ab\rightarrow cd}_{\tilde{g}}$ appears to have no relation to $H^{ab\rightarrow dc}_{\tilde{g}}$ might seem a bit strange.  However, one can see this will be the case, e.g., by noticing the $k_{\perp}$ dependence changes when one interchanges final state partons.

We conclude this section with a brief discussion of a future numerical study and of the key insights a measurement of this observable at RHIC might provide.  In order to estimate the size of $A_{LT}$ for hadron, jet, and photon production, we must determine the input for the twist-3 functions that enter into (\ref{e:sigmahadron}).  We can obtain information on the function $\tilde{g}(x)$ through its relation (\ref{e:g1T}) to $g_{1T}(x,\vec{k}_{\perp}^{2})$ --- see \cite{Kang:2011jw} for a recent example as well as \cite{Metz:2008ib, Kotzinian:2006dw, Avakian:2007mv}.  In addition, one can choose to pull the $\xi$-independent terms in the hard factors for $H_{F_{DT}}^{i}$ and $H_{G_{DT}}^{i}$ out of the integral over $x_{1}$ and write a term involving $g_{T}(x)$ --- see the discussion in the previous paragraph.  We also have information on this function, e.g., through the Wandzura-Wliczek approximation \cite{Wandzura:1977qf, Accardi:2009au}.  The main obstacle then is the $qgq$ correlator contributions.  The off-diagonal contributions (i.e., $x\neq x_{1}$) to $F_{FT}$ and $G_{FT}$ needed for this DSA observable are not as well-determined as the diagonal pieces that enter into SSAs.  In Ref.~\cite{Kang:2008ey}, a Gaussian form was assumed for $F_{FT}(x,x_{1})$ ($T_{q,F}$ in their notation) that was a maximum for $x=x_{1}$ and fell off for $x\neq x_{1}$.  This study was done in the context of the evolution of $F_{FT}(x,x)$.  In Ref.~\cite{Braun:2011aw}, an analysis of higher-twist functions was conducted using light-cone wave functions that include $qqqg$ Fock states.  In contrast to \cite{Kang:2008ey}, this study found $F_{FT}(x,x_{1})$ ($T_{qFq}$ in their notation) takes on its greatest values when $x\neq x_{1}$ and some of its lowest values when $x=x_{1}$.  It is our plan to determine the impact of the $qgq$ correlators on the size of $A_{LT}$ and provide a complete estimate for the observable.

Given this estimate, the importance of the measurement of $A_{LT}$ in pion production from proton-proton ($pp$) collisions at RHIC is threefold.  First, through this observable one might be able to probe the gluon helicity $\Delta g$ down to momentum fractions $x\sim10^{-3}$ (or even lower), more than an order of magnitude below the sensitivity of all current measurements \cite{Abelev:2006uq, Adamczyk:2012qj, Adler:2004ps, Adare:2007dg, Ageev:2005pq, Alekseev:2008cz, Adolph:2012vj, Airapetian:1999ib, Adeva:2004dh}.  Given the recent debate as to the size of $\Delta g$ at smaller $x$ --- see \cite{deFlorian:2011ia} and references therein, $A_{LT}$ could offer valuable insight into the matter.  Second, a measurement of $A_{LT}$ in this process would be a first step towards extracting (non-diagonal) information on the 3-parton correlators $F_{FT}$ and $G_{FT}$.  Information on these functions is beneficial in its own right, but, as alluded to in the previous paragraph, one must know the off-diagonal contributions to $F_{FT}$ and $G_{FT}$ in order to fully determine the evolution of $F_{FT}(x,x)$ \cite{Kang:2008ey, Braun:2009mi, Vogelsang:2009pj, Zhou:2008mz, Ma:2012xn}.  This evolution is a vital aspect if one wants to fully understand SSAs.  Finally, the ``sign mismatch'' issue that has arisen involving $F_{FT}(x,x)$ (or $T_{F}(x,x)$) and the Sivers function \cite{Kang:2011hk} has called into question whether the collinear twist-3 framework is the correct formalism to describe, e.g., the large SSAs seen in inclusive hadron production from $pp$ collisions \cite{Bunce:1976yb, Adams:1991rw, Krueger:1998hz, Adams:2003fx, Adler:2005in, :2008mi, Adamczyk:2012xd}.  The study of DSAs may provide new insights on this point.  For instance, should a significant discrepancy exist between a numerical estimate of $A_{LT}$ in pion production and a future measurement of this observable at RHIC, one may ask whether the collinear twist-3 approach taken in this paper is the appropriate mechanism to consider for both SSAs and DSAs.  Of course, one must keep in mind that knowledge of the relevant twist-3 functions at present is rather limited \cite{Kang:2008ey, Braun:2011aw}.

%
%
%
\section{Summary}
\label{s:sum}
In conclusion, we have calculated the double-spin dependent cross section for inclusive hadron and jet production for the longitudinal-transverse asymmetry $A_{LT}$ in nucleon-nucleon collisions within the collinear twist-3 framework.  We have also reviewed the results for $A_{LT}$ in direct photon production from nucleon-nucleon scattering \cite{Liang:2012rb}.  These derivations are the DSA analog to the SSAs calculated in the same processes \cite{Qiu:1991pp, Qiu:1998ia, Kouvaris:2006zy, Koike:2007rq}.  Furthermore, these reactions require a complete set of collinear twist-3 functions for a transversely polarized nucleon in order to fully describe the process.  We emphasize again that we did not consider contributions involving chiral-odd correlation functions.  We have found that the solution, when written in terms of D-type functions, allows for a ``compact'' form involving $\tilde{g}(x)$ and its derivative; similar forms have manifested themselves in other reactions \cite{Qiu:1991pp, Kouvaris:2006zy, Koike:2007rq, Metz:2010xs, Kang:2011jw, Liang:2012rb}.  We also made some intriguing observations on the structure of the hard factors, in particular for $H_{F_{DT}}^{i}$ and $H_{G_{DT}}^{i}$.

In addition, we have briefly outlined our plan for a future numerical study of this observable.  The main difficulty underlying such an analysis is how to handle the contributions from the 3-parton correlators $F_{FT}$ and $G_{FT}$ since, unlike the case for SSAs, $A_{LT}$ is sensitive to the off-diagonal contributions to these functions.  Such an undertaking is worthwhile, however, since a measurement of this effect at RHIC could provide insight on some important areas of research in hadronic spin physics.  These include not only information on 3-parton correlators, which are important in their own right, but also access to the gluon helicity distribution $\Delta g$ at momentum fractions not yet explored ($x\sim10^{-3}$), information on the evolution of the ETQS function $T_{F}(x,x)$ that appears in SSAs, and a general understanding of the mechanism that causes twist-3 spin asymmetries in nucleon-nucleon collisions. 
\\[0.5cm]
%
%
\noindent
{\bf Acknowledgments:}
We would like to thank C.~Aidala, E.~Aschenauer, and B.~Surrow for helpful exchanges on the possibility of measuring this observable at RHIC.  We also appreciate useful discussions with V.~Braun and A.~Manashov with regards to Ref.~\cite{Braun:2011aw}.  This work has been supported by the NSF under Grant No.~PHY-1205942 and by the BMBF under Grant No.~OR 06RY9191.

\begin{appendices}
\appendixpage
%
%
%
%
\section{Hard scattering coefficients for hadron production} \label{a:hadron}
Here we give the hard scattering coefficients $H^{i}$ for hadron production.  Table \ref{t:1} lists all the channels $i$ ($ab\rightarrow cd$) and gives the corresponding partonic Mandelstam variable $\hat{m}_{i}$ for that channel.  Note that $(\hat{t}\leftrightarrow\hat{u})$ means interchange of $\hat{t}$ and $\hat{u}$.  We define $\xi = x_{g}/x$, where $x_{g} = x-x_1$, and understand $1/\xi$ to mean $PV(1/\xi)$.  We also mention that the $SU(3)$ color factors depend on $N_{c}=3$.  The double-spin dependent cross section for jet production takes on the same form as Eq.~(\ref{e:sigmahadron}) but now with $D_{1}^{C/c}(z) = \delta(1-z)$.  This allows hard factors to be combined for channels that differ by an interchange of the final state partons.\vspace{5cm}
\\[0.3cm]
\begin{table}[h] 
\centering
\begin{tabular} {c || c | c | c}
$\hat{m}_{i}$  & $\hat{s}$ & $\hat{t}$ & $\hat{u}$ \\
\hline
$i\,(ab\rightarrow cd)$& $q\bar{q}\rightarrow gg$ & $qg\rightarrow gq$ & $qg\rightarrow qg$ \\
& $qq\rightarrow qq$  & $qq'\rightarrow q'q$ & $qq'\rightarrow qq'$ \\
& & $q\bar{q}\rightarrow \bar{q}q$ & $q\bar{q}\rightarrow q\bar{q}$ \\
& & $q\bar{q}' \rightarrow \bar{q}'q$ & $q\bar{q}' \rightarrow q\bar{q}'$ \\
& & \,$q\bar{q} \rightarrow \bar{q}'q$ & \,\,\,$q\bar{q} \rightarrow q'\bar{q}'$
\end{tabular}
\caption{Mandelstam variable $\hat{m}_{i}$ for each channel $i\,(ab\rightarrow cd)$.} \label{t:1}
\end{table}\\[0.5cm] 
\underline{$qg\rightarrow qg$ channel}
\begin{align}
H_{\tilde{g}}&= \frac{1} {2} \left[\frac{(\hat{s}-\hat{u})\hat{u}} {\hat{s}\hat{t}}\right] + \frac{1} {2N_{c}^{2}}\left[\frac{\hat{s}-\hat{u}} {\hat{u}}\right]+\frac{1} {2(N_{c}^{2}-1)}\left[\frac{(\hat{s}-\hat{u})^{2}} {\hat{t}^{2}}\right]\\[0.3cm]
H_{G_{DT}}&= \frac{1} {2}\left[\frac{\hat{s}(\hat{s}^{2}-\hat{t}\hat{u})} {\hat{t}^{2}\hat{u}} - \frac{\hat{u}^{2}(\hat{t}-\hat{u})} {\hat{s}\hat{t}^{2}} - \frac{(\hat{s}^{2}+\hat{u}^{2})(\hat{t}^{2}-3\hat{s}\hat{u})} {(1-\xi)\hat{s}\hat{t}^{2}\hat{u}} + \frac{2\hat{s}(\hat{s}-\hat{u})} {\xi \hat{t}\hat{u}}\right]\nonumber \\[0.1cm]
&\hspace{0.4cm} +\,\frac{1} {2N_{c}^{2}}\left[\frac{1} {1-\xi}-\frac{\hat{s}^{2}+2\hat{u}^{2}} {\hat{s}\hat{u}}+\frac{2(\hat{s}-\hat{u})} {\xi\hat{s}}\right] - \frac{1} {2(N_{c}^{2}-1)}\left[\frac{(\hat{s}-\hat{u})^{2}} {\hat{t}^{2}}\left(-\frac{1} {1-\xi}-\frac{2} {\xi}\right)\right]\\[0.3cm]
H_{F_{DT}}&=  \frac{1} {2}\left[\frac{\hat{s}(\hat{s}^{2}-\hat{t}\hat{u})} {\hat{t}^{2}\hat{u}} - \frac{\hat{u}^{2}(\hat{t}-\hat{u})} {\hat{s}\hat{t}^{2}} + \frac{(\hat{s}^{2}+\hat{u}^{2})(\hat{t}^{2}-3\hat{s}\hat{u})} {(1-\xi)\hat{s}\hat{t}^{2}\hat{u}}\right]\nonumber \\[0.1cm]
&\hspace{0.4cm} +\,\frac{1} {2N_{c}^{2}}\left[-\frac{1} {1-\xi} -\frac{\hat{s}^{2}+2\hat{u}^{2}} {\hat{s}\hat{u}}\right] - \frac{1} {2(N_{c}^{2}-1)}\left[\frac{(\hat{s}-\hat{u})^{2}} {(1-\xi)\hat{t}^{2}}\right]
\end{align}
\\[0.5cm]
\underline{$qg\rightarrow gq$ channel}
\begin{align}
H_{\tilde{g}}&= \frac{1} {2}\left[\frac{(\hat{s}-\hat{t})^{2}(\hat{t}\hat{u}-\hat{s}^{2})} {\hat{s}\hat{t}\hat{u}^{2}}\right]+\frac{1} {2(N_{c}^{2}-1)}\left[\frac{(\hat{s}-\hat{t})^{2}} {\hat{u}^{2}}\right]\\[0.3cm]
H_{G_{DT}}&= -H_{G_{DT}}^{qg\rightarrow qg}(\hat{t}\leftrightarrow\hat{u})\\[0.3cm]
H_{F_{DT}}&= -H_{F_{DT}}^{qg\rightarrow qg}(\hat{t}\leftrightarrow\hat{u})
\end{align}
\\[0.5cm]
\underline{$q\bar{q}\rightarrow gg$ channel}
\begin{align}
H_{\tilde{g}}&= -\frac{1} {2N_{c}}\left[\frac{(\hat{t}^{2}+\hat{u}^{2})(\hat{s}^{3}+2\hat{t}^{2}\hat{u})} {\hat{s}\hat{t}^{2}\hat{u}^{2}}\right]+\frac{N_{c}} {2}\left[\frac{\hat{s}^{2}-2\hat{t}\hat{u}} {\hat{t}^{2}}\right]-\frac{1} {2N_{c}^{3}}\left[\frac{\hat{s}(\hat{t}^{2}+\hat{u}^{2})} {\hat{u}^{2}\hat{t}}\right]\\[0.3cm]
H_{G_{DT}}&= \frac{1} {2N_{c}}\left[\frac{\hat{s}(\hat{t}-\hat{u})(\hat{s}^{2}-\hat{t}\hat{u})} {(1-\xi)\hat{t}^{2}\hat{u}^{2}} + \frac{2\hat{s}(\hat{t}-\hat{u})(\hat{t}^{2}+\hat{u}^{2})} {\xi\hat{t}^{2}\hat{u}^{2}} - \frac{(\hat{t}-\hat{u})} {\hat{s}} \left(\frac{2\hat{s}^{4}} {\hat{t}^{2}\hat{u}^{2}} + \frac{(\hat{t}-\hat{u})^{2}} {\hat{s}\hat{t}\hat{u}}\right)\right]\nonumber \\[0.1cm]
&\hspace{0.4cm} -\,\frac{N_{c}} {2}\left[\frac{(\hat{u}-\hat{t})(\hat{t}^{2}+\hat{u}^{2})(\hat{s}^{2}-\hat{t}\hat{u})} {\hat{s}\hat{t}^{2}\hat{u}^{2}}\left(-\frac{1} {1-\xi}-\frac{2} {\xi}\right)-\frac{(\hat{t}-\hat{u})(\hat{t}^{2}+\hat{u}^{2})(\hat{s}^{2}+2\hat{t}\hat{u})} {\hat{s}\hat{t}^{2}\hat{u}^{2}}\right] \nonumber \\[0.1cm]
&\hspace{0.4cm} +\, \frac{1} {2N_{c}^{3}}\left[-\frac{(\hat{t}-\hat{u})\hat{s}} {(1-\xi)\hat{t}\hat{u}} + \frac{\hat{s}(\hat{t}-\hat{u})(\hat{s}^{2}+\hat{u}\hat{t})} {\hat{t}^{2}\hat{u}^{2}}\right]  \\[0.3cm]
H_{F_{DT}}&= \frac{1} {2N_{c}}\left[-\frac{\hat{s}(\hat{t}-\hat{u})(\hat{s}^{2}-\hat{t}\hat{u})} {(1-\xi)\hat{t}^{2}\hat{u}^{2}} - \frac{(\hat{t}-\hat{u})} {\hat{s}} \left(\frac{2\hat{s}^{4}} {\hat{t}^{2}\hat{u}^{2}} + \frac{(\hat{t}-\hat{u})^{2}} {\hat{s}\hat{t}\hat{u}}\right)\right]\nonumber \\[0.1cm]
&\hspace{0.4cm} -\,\frac{N_{c}} {2}\left[\frac{(\hat{u}-\hat{t})(\hat{t}^{2}+\hat{u}^{2})(\hat{s}^{2}-\hat{t}\hat{u})} {(1-\xi)\hat{s}\hat{t}^{2}\hat{u}^{2}}-\frac{(\hat{t}-\hat{u})(\hat{t}^{2}+\hat{u}^{2})(\hat{s}^{2}+2\hat{t}\hat{u})} {\hat{s}\hat{t}^{2}\hat{u}^{2}}\right] \nonumber \\[0.1cm]
&\hspace{0.4cm} +\, \frac{1} {2N_{c}^{3}}\left[\frac{(\hat{t}-\hat{u})\hat{s}} {(1-\xi)\hat{t}\hat{u}}+\frac{\hat{s}(\hat{t}-\hat{u})(\hat{s}^{2}+\hat{u}\hat{t})} {\hat{t}^{2}\hat{u}^{2}}\right] 
\end{align}
\\[0.5cm]
\underline{$qq'\rightarrow qq'$ channel}
\begin{align}
H_{\tilde{g}}&=  -\frac{1} {2N_{c}^{2}}\left[\frac{(\hat{t}-\hat{u})(\hat{s}-\hat{u})} {\hat{t}^{2}}\right]\\[0.3cm]
H_{G_{DT}}&=  -\frac{1} {2}\left[\frac{2(\hat{s}-\hat{u})} {\xi\hat{t}}-\frac{\hat{s}(\hat{s}-\hat{u})} {(1-\xi)\hat{t}^{2}}-\frac{\hat{u}} {\hat{t}}\right] + \frac{1} {2N_{c}^{2}}\left[\frac{2(\hat{s}-\hat{u})} {\hat{t}}\left(\frac{2\hat{t}-\hat{u}} {\xi\hat{t}}+\frac{1} {1-\xi}\right)-\frac{\hat{u}} {\hat{t}}\right]   \\[0.3cm]
H_{F_{DT}}&= -\frac{1} {2}\left[\frac{\hat{s}(\hat{s}-\hat{u})} {(1-\xi)\hat{t}^{2}}-\frac{\hat{u}} {\hat{t}}\right]+\frac{1} {2N_{c}^{2}}\left[-\frac{2(\hat{s}-\hat{u})} {\hat{t}(1-\xi)}-\frac{\hat{u}} {\hat{t}}\right]
\end{align}
\\[0.5cm]
\underline{$qq'\rightarrow q'q$ channel}
\begin{align}
H_{\tilde{g}}&=  -\frac{1} {2} \left[\frac{\hat{s}(\hat{t}-\hat{s})} {\hat{u}^{2}}\right]+\frac{1} {2N_{c}^{2}}\left[\frac{2(\hat{s}-\hat{t})} {\hat{u}}\right]     \\[0.3cm]
H_{G_{DT}}&= -H_{G_{DT}}^{qq'\rightarrow qq'}(\hat{t}\leftrightarrow\hat{u})\\[0.3cm]
H_{F_{DT}}&= -H_{F_{DT}}^{qq'\rightarrow qq'}(\hat{t}\leftrightarrow\hat{u})
\end{align}
\\[0.5cm]
\underline{$qq\rightarrow qq$ channel}
\begin{align}
H_{\tilde{g}}&= \frac{1} {2}\left[\frac{\hat{s}^{2}(\hat{s}-\hat{t})} {\hat{t}\hat{u}^{2}}\right]-\frac{1} {2N_{c}^{2}}\left[\frac{3\hat{s}} {\hat{u}}-\frac{\hat{s}(\hat{s}-\hat{u})} {\hat{t}^{2}}\right]-\frac{1} {2N_{c}}\left[\frac{2\hat{s}^{3}} {\hat{u}^{2}\hat{t}}\right]+\frac{1} {2N_{c}^{3}}\left[\frac{2\hat{s}^{3}} {\hat{u}\hat{t}^{2}}\right]     \\[0.3cm]
H_{G_{DT}}&= \frac{1} {2}\left[\frac{2\hat{s}^{3}(\hat{u}-\hat{t})} {(1-\xi)\hat{t}^{2}\hat{u}^{2}}+\frac{2\hat{s}(\hat{u}-\hat{t})} {\xi\hat{t}\hat{u}}+\frac{\hat{s}(\hat{u}-\hat{t})} {\hat{t}\hat{u}}\right]\nonumber \\[0.1cm]
&\hspace{0.4cm}-\,\frac{1} {2N_{c}^{2}}\left[\frac{2\hat{s}(\hat{t}-\hat{u})(2\hat{s}^{2}-3\hat{t}\hat{u})} {\xi\hat{t}^{2}\hat{u}^{2}}-\frac{\hat{s}(\hat{t}-\hat{u})} {\hat{t}\hat{u}}\left(1+\frac{2} {1-\xi}\right)\right]\nonumber \\[0.1cm]
&\hspace{0.4cm} -\,\frac{1} {2N_{c}}\left[\frac{\hat{s}^{3}(\hat{u}-\hat{t})} {\hat{t}^{2}\hat{u}^{2}}\left(1+\frac{1} {1-\xi}\right)\right]+\frac{1} {2N_{c}^{3}}\left[\frac{\hat{s}^{3}(\hat{t}-\hat{u})} {\hat{t}^{2}\hat{u}^{2}}\left(-1+\frac{1} {1-\xi}+\frac{4} {\xi}\right)\right]   \\[0.3cm]
H_{F_{DT}}&=  \frac{1} {2}\left[-\frac{2\hat{s}^{3}(\hat{u}-\hat{t})} {(1-\xi)\hat{t}^{2}\hat{u}^{2}}+\frac{\hat{s}(\hat{u}-\hat{t})} {\hat{t}\hat{u}}\right]-\frac{1} {2N_{c}^{2}}\left[-\frac{\hat{s}(\hat{t}-\hat{u})} {\hat{t}\hat{u}}\left(1-\frac{2} {1-\xi}\right)\right] \nonumber \\[0.1cm]
&\hspace{0.4cm} -\, \frac{1} {2N_{c}}\left[\frac{\hat{s}^{3}(\hat{u}-\hat{t})} {\hat{t}^{2}\hat{u}^{2}}\left(1-\frac{1} {1-\xi}\right)\right] + \frac{1} {2N_{c}^{3}}\left[\frac{\hat{s}^{3}(\hat{t}-\hat{u})} {\hat{t}^{2}\hat{u}^{2}}\left(-1-\frac{1} {1-\xi}\right)\right] 
\end{align}
\\[0.5cm]
\underline{$q\bar{q}\rightarrow q\bar{q}$ channel}
\begin{align}
H_{\tilde{g}}&= -\frac{1} {2}\left[\frac{\hat{u}(\hat{s}-2\hat{t})} {\hat{s}\hat{t}}\right]-\frac{1} {2N_{c}^{2}}\left[\frac{\hat{u}^{2}} {\hat{s}\hat{t}}\left(\frac{2\hat{s}^{2}+9\hat{t}^{2}} {\hat{u}\hat{t}}+\frac{\hat{s}-2\hat{u}} {\hat{s}}\right)\right] - \frac{1} {2N_{c}^{3}}\left[\frac{2\hat{u}^{2}(\hat{s}-\hat{t})} {\hat{s}\hat{t}^{2}}\right]    \\[0.3cm]
H_{G_{DT}}&=  \frac{1} {2}\left[\frac{\hat{u}(3\hat{t}^{2}+\hat{u}^{2})} {\hat{s}^{2}\hat{t}} + \frac{2\hat{u}(\hat{t}^{2}+\hat{u}^{2})} {\xi\hat{s}^{2}\hat{t}}-\frac{2\hat{u}(\hat{s}^{3}+\hat{u}\hat{t}^{2})} {(1-\xi)\hat{s}^{2}\hat{t}^{2}}\right] \nonumber \\[0.1cm]
&\hspace{0.4cm} -\,\frac{1} {2N_{c}^{2}}\left[\frac{\hat{u}(3\hat{t}^{2}+\hat{u}^{2})} {\hat{s}^{2}\hat{t}}-\frac{2\hat{u}(\hat{s}-2\hat{t})} {(1-\xi)\hat{s}\hat{t}}+\frac{2\hat{u}(2\hat{s}^{2}-\hat{t}(\hat{s}-4\hat{t}))} {\xi\hat{s}\hat{t}^{2}}\right]  \nonumber \\[0.1cm]
&\hspace{0.4cm} -\, \frac{1} {2N_{c}}\left[-\frac{2\hat{u}^{2}} {\hat{s}\hat{t}}\left(2+\frac{1} {1-\xi}\right)-\frac{\hat{u}^{3}} {\hat{s}\hat{t}^{2}}\left(1+\frac{1} {1-\xi}\right)-\frac{4\hat{u}^{2}} {\xi\hat{s}\hat{t}}\right] \nonumber \\[0.1cm]
&\hspace{0.4cm}+\, \frac{1} {2N_{c}^{3}}\left[-\frac{\hat{u}^{2}(\hat{t}-\hat{s})} {\hat{s}\hat{t}^{2}}\left(1-\frac{1} {1-\xi}\right) -\frac{2\hat{u}^{2}} {\hat{s}\hat{t}}-\frac{4\hat{u}^{2}} {\xi\hat{t}^{2}}\right]
\\[0.3cm]
H_{F_{DT}}&=  \frac{1} {2}\left[\frac{\hat{u}(3\hat{t}^{2}+\hat{u}^{2})} {\hat{s}^{2}\hat{t}}+\frac{2\hat{u}(\hat{s}^{3}+\hat{u}\hat{t}^{2})} {(1-\xi)\hat{s}^{2}\hat{t}^{2}}\right] -\frac{1} {2N_{c}^{2}}\left[\frac{\hat{u}(3\hat{t}^{2}+\hat{u}^{2})} {\hat{s}^{2}\hat{t}}+\frac{2\hat{u}(\hat{s}-2\hat{t})} {(1-\xi)\hat{s}\hat{t}}\right]  \nonumber \\[0.1cm]
&\hspace{0.4cm} -\, \frac{1} {2N_{c}}\left[-\frac{2\hat{u}^{2}} {\hat{s}\hat{t}}\left(2-\frac{1} {1-\xi}\right)-\frac{\hat{u}^{3}} {\hat{s}\hat{t}^{2}}\left(1-\frac{1} {1-\xi}\right)\right] \nonumber\\[0.1cm]
&\hspace{0.4cm}+\, \frac{1} {2N_{c}^{3}}\left[-\frac{\hat{u}^{2}(\hat{t}-\hat{s})} {\hat{s}\hat{t}^{2}}\left(1+\frac{1} {1-\xi}\right) -\frac{2\hat{u}^{2}} {\hat{s}\hat{t}}\right]
\end{align}
\\[0.5cm]
\underline{$q\bar{q}\rightarrow \bar{q}q$ channel}
\begin{align}
H_{\tilde{g}}&=  \frac{1} {2}\left[\frac{\hat{t}(\hat{t}-\hat{s})} {\hat{u}^{2}}\right]+\frac{1} {2N_{c}^{2}}\left[\frac{3\hat{t}} {\hat{u}}-\frac{2\hat{t}(\hat{s}-\hat{u})} {\hat{s}^{2}}\right]-\frac{1}{2N_{c}}\left[\frac{2\hat{t}^{2}} {\hat{u}^{2}}\right]+\frac{1} {2N_{c}^{3}}\left[\frac{2\hat{t}^{2}} {\hat{s}\hat{u}}\right]     \\[0.3cm]
H_{G_{DT}}&= -H_{G_{DT}}^{q\bar{q}\rightarrow q\bar{q}}(\hat{t}\leftrightarrow\hat{u})\\[0.3cm]
H_{F_{DT}}&= -H_{F_{DT}}^{q\bar{q}\rightarrow q\bar{q}}(\hat{t}\leftrightarrow\hat{u})
\end{align}
\\[0.5cm]
\underline{$q\bar{q}\rightarrow q'\bar{q}'$ channel}
\begin{align}
H_{\tilde{g}}&= \frac{1} {2}\left[\frac{\hat{t}^{2}+\hat{u}^{2}} {\hat{s}\hat{t}}\right]+\frac{1} {2N_{c}^{2}}\left[\frac{(\hat{t}-2\hat{s})(\hat{t}^{2}+\hat{u}^{2})} {\hat{s}^{2}\hat{t}}\right]   \\[0.3cm]
H_{G_{DT}}&= \frac{1} {2}\left[\frac{2\hat{u}(\hat{t}^{2}+\hat{u}^{2})} {\xi\hat{s}^{2}\hat{t}} +\frac{2\hat{u}(\hat{t}-\hat{u})} {\hat{s}^{2}} +\frac{\hat{u}(\hat{t}^{2}+\hat{u}^{2})} {(1-\xi)\hat{s}^{2}\hat{t}}\right] + \frac{1} {2N_{c}^{2}}\left[\frac{2(\hat{t}^{2}+\hat{u}^{2})} {\hat{s}\hat{t}}\left(\frac{1} {1-\xi}+\frac{2} {\xi}\right)-\frac{2\hat{u}(\hat{t}-\hat{u})} {\hat{s}^{2}}\right]  \\[0.3cm]
H_{F_{DT}}&=  \frac{1} {2}\left[\frac{2\hat{u}(\hat{t}-\hat{u})} {\hat{s}^{2}} -\frac{\hat{u}(\hat{t}^{2}-\hat{u}^{2})} {(1-\xi)\hat{s}^{2}\hat{t}}\right] + \frac{1} {2N_{c}^{2}}\left[-\frac{2\hat{u}(\hat{t}-\hat{u})} {\hat{s}^{2}}-\frac{2(\hat{t}^{2}+\hat{u}^{2})} {(1-\xi)\hat{s}\hat{t}}\right] 
\end{align}
\\[0.5cm]
\underline{$q\bar{q}\rightarrow \bar{q}'q'$ channel}
\begin{align}
H_{\tilde{g}}&= \frac{1} {2N_{c}^{2}}\left[\frac{(\hat{s}-\hat{u})(\hat{t}^{2}+\hat{u}^{2})} {\hat{s}^{2}\hat{u}}\right]   \\[0.3cm]
H_{G_{DT}}&= -H_{G_{DT}}^{q\bar{q}\rightarrow q'\bar{q}'}(\hat{t}\leftrightarrow\hat{u})\\[0.3cm]
H_{F_{DT}}&= -H_{F_{DT}}^{q\bar{q}\rightarrow q'\bar{q}'}(\hat{t}\leftrightarrow\hat{u})
\end{align}
\\[0.5cm]
\underline{$q\bar{q}'\rightarrow q\bar{q}'$ channel}
\begin{align}
H_{\tilde{g}}&= \frac{1} {2}\left[\frac{\hat{u}-\hat{s}} {\hat{t}}\right]+\frac{1} {2N_{c}^{2}}\left[\frac{(\hat{s}-2\hat{t})(\hat{u}-\hat{s})} {\hat{t}^{2}}\right]   \\[0.3cm]
H_{G_{DT}}&= \frac{1} {2} \left[\frac{\hat{u}(\hat{s}-\hat{u})} {(1-\xi)\hat{t}^{2}} - \frac{\hat{u}} {\hat{t}}\right]+\frac{1} {2N_{c}^{2}}\left[\frac{2(\hat{s}-\hat{t})(\hat{s}-\hat{u})} {\xi\hat{t}^{2}}-\frac{2(\hat{s}-\hat{u})} {(1-\xi)\hat{t}} -\frac{\hat{u}} {\hat{t}}\right]\\[0.3cm]
H_{F_{DT}}&= \frac{1} {2} \left[-\frac{\hat{u}(\hat{s}-\hat{u})} {(1-\xi)\hat{t}^{2}} - \frac{\hat{u}} {\hat{t}}\right]+\frac{1} {2N_{c}^{2}}\left[\frac{2(\hat{s}-\hat{u})} {(1-\xi)\hat{t}} -\frac{\hat{u}} {\hat{t}}\right]
\end{align}
\\[0.5cm]
\underline{$q\bar{q}'\rightarrow \bar{q}'q$ channel}
\begin{align}
H_{\tilde{g}}&= \frac{1} {2} \left[\frac{\hat{t}(\hat{t}-\hat{s})} {\hat{u}^{2}}\right]-\frac{1} {2N_{c}^{2}}\left[\frac{2(\hat{s}-\hat{t})} {\hat{u}}\right]  \\[0.3cm]
H_{G_{DT}}&= -H_{G_{DT}}^{q\bar{q}'\rightarrow q\bar{q}'}(\hat{t}\leftrightarrow\hat{u})\\[0.3cm]
H_{F_{DT}}&= -H_{F_{DT}}^{q\bar{q}'\rightarrow q\bar{q}'}(\hat{t}\leftrightarrow\hat{u})
\end{align}
\\[1cm]
%
%
%
%
\section{Hard scattering coefficients for photon production} \label{a:photon}
Here we give the hard scattering coefficients $H^{i}$ for direct photon production.  We define $\xi = x_{g}/x$, where $x_g = x-x_1$, and understand $1/\xi$ to mean $PV(1/\xi)$.  We also note that the $SU(3)$ color factors depend on $C_{F} =4/3$ and $N_{c}=3$. The double-spin dependent cross section has the same form as Eq.~(\ref{e:sigmahadron}), but now we set $D_{1}^{C/c}(z) = \delta(1-z)$ and for one factor of $\alpha_{s}$ make the replacement $\alpha_{s}\rightarrow \alpha_{em} e_{a}^{2}$, where $e_{a}$ is the charge of a quark with flavor $a$ (in units of $e$).
\\[0.5cm]
\underline{$qg\rightarrow \gamma q$ channel}
\begin{align}
H_{\tilde{g}}&= \frac{N_c}{N_c^2 -1}
\bigg[ \frac{\hat{t}^2 - \hat{s}^2}{\hat{s}\hat{t}} \bigg] 
\nonumber \\
H_{G_{DT}}&= \frac{N_c}{N_c^2-1} 
\bigg[ \frac{(\hat{t}^2 - \hat{s}^2)}{\hat{s}\hat{t}}\left(-\frac{1} {1-\xi}-\frac{2} {\xi}\right) \bigg]
- \frac{1}{N_c} \bigg[ \frac{\hat{u} \, (\hat{s}^2 + 2\hat{t}^2)}{\hat{s} \hat{t}^2} 
+ \frac{\hat{u}}{\hat{t}}\left(-\frac{1} {1-\xi}-\frac{2} {\xi}\right) 
+ \frac{2 \hat{u}}{\xi \, \hat{s}} \bigg] 
\nonumber \\
H_{F_{DT}}&= \frac{N_c}{N_c^2-1} 
\bigg[ \frac{\hat{t}^2 - \hat{s}^2}{(1 - \xi) \, \hat{s} \, \hat{t}} \bigg]
- \frac{1}{N_c} \bigg[ \frac{\hat{u} \, (\hat{s}^2 + 2\hat{t}^2)}{\hat{s} \, \hat{t}^2} 
+ \frac{\hat{u}}{(1-\xi) \, \hat{t}} \bigg] 
\end{align}
\\[0.5cm]
\underline{$q\bar{q}\rightarrow \gamma g$ channel}
\begin{align}
H_{\tilde{g}}&=  \frac{1}{N_c^2 }
\bigg[ \frac{\hat{t}^2 + \hat{u}^2}{\hat{t} \hat{u}} \bigg] 
\nonumber \\
H_{G_{DT}}&=
 \frac{(\hat{t}^2 + \hat{u}^2)}{ \hat{t} \hat{u}} \left(-\frac{1} {1-\xi}-\frac{2} {\xi}\right) 
+ \frac{2C_F}{N_c} \bigg[ \frac{\hat{s}^2 \, (\hat{t} - \hat{u})}{\hat{t}^2 \hat{u}}
- \frac{(\hat{t} - \hat{u})}{ \hat{t}}\left(-1+\frac{1} {1-\xi}\right)  \bigg] 
\nonumber \\
H_{F_{DT}} &=
 \frac{\hat{t}^2+\hat{u}^2}{(1 - \xi)\hat{t}\hat{u}} 
+\frac{2C_F}{N_c} \bigg[ \frac{\hat{s}^2 \, (\hat{t} - \hat{u})}{\hat{t}^2 \hat{u}} 
- \frac{(\hat{t} - \hat{u})}{\hat{t}}\left(-1-\frac{1} {1-\xi}\right) \bigg] 
\end{align}

\end{appendices}


\begin{thebibliography}{99}

\bibitem{Bunce:1976yb} 
  G.~Bunce {\it et al.},
  Phys.\ Rev.\ Lett.\  {\bf 36}, 1113 (1976).

\bibitem{Adams:1991rw} 
  D.~L.~Adams {\it et al.} [E581 and E704 Collaborations],
  Phys.\ Lett.\ B {\bf 261}, 201 (1991);
  D.~L.~Adams {\it et al.} [E704 Collaboration],
  Phys.\ Lett.\ B {\bf 264}, 462 (1991).

\bibitem{Krueger:1998hz} 
  K.~Krueger {\it et al.},
  Phys.\ Lett.\ B {\bf 459}, 412 (1999).

\bibitem{Adams:2003fx}
  J.~Adams {\it et al.} [STAR Collaboration],
  Phys.\ Rev.\ Lett.\  {\bf 92}, 171801 (2004)
  [hep-ex/0310058];
  B.~I.~Abelev {\it et al.} [STAR Collaboration],
  Phys.\ Rev.\ Lett.\  {\bf 101}, 222001 (2008)
  [arXiv:0801.2990 [hep-ex]].

\bibitem{Adler:2005in} 
  S.~S.~Adler {\it et al.}  [PHENIX Collaboration],
  Phys.\ Rev.\ Lett.\  {\bf 95}, 202001 (2005)
  [hep-ex/0507073].

\bibitem{:2008mi} 
  I.~Arsene {\it et al.}  [BRAHMS Collaboration],
  Phys.\ Rev.\ Lett.\  {\bf 101}, 042001 (2008)
  [arXiv:0801.1078 [nucl-ex]].
  
\bibitem{Adamczyk:2012xd} 
  L.~Adamczyk {\it et al.}  [STAR Collaboration],
  Phys.\ Rev.\ D {\bf 86}, 051101 (2012)
  [arXiv:1205.6826 [nucl-ex]].
  
\bibitem{Kane:1978nd} 
  G.~L.~Kane, J.~Pumplin and W.~Repko,
  Phys.\ Rev.\ Lett.\  {\bf 41}, 1689 (1978).

\bibitem{Ma:2008gm}
  J.~P.~Ma and H.~Z.~Sang,
  JHEP {\bf 0811}, 090 (2008)
  [arXiv:0809.4811 [hep-ph]];
  Phys.\ Lett.\  B {\bf 676}, 74 (2009)
  [arXiv:0811.0224 [hep-ph]].
  
\bibitem{Efremov:1981sh}
  A.~V.~Efremov and O.~V.~Teryaev,
  Sov.\ J.\ Nucl.\ Phys.\  {\bf 36}, 140 (1982)
  [Yad.\ Fiz.\  {\bf 36}, 242 (1982)];
  Phys.\ Lett.\ B {\bf 150}, 383 (1985).

\bibitem{Qiu:1991pp} 
  J.-w.~Qiu and G.~F.~Sterman,
  Phys.\ Rev.\ Lett.\  {\bf 67}, 2264 (1991);
  Nucl.\ Phys.\ B {\bf 378}, 52 (1992).

\bibitem{Qiu:1998ia} 
  J.-w.~Qiu and G.~F.~Sterman,
  Phys.\ Rev.\ D {\bf 59}, 014004 (1999)
  [hep-ph/9806356].

\bibitem{Eguchi:2006qz}
  H.~Eguchi, Y.~Koike and K.~Tanaka,
  Nucl.\ Phys.\  B {\bf 752}, 1 (2006)
  [arXiv:hep-ph/0604003];
  Nucl.\ Phys.\  B {\bf 763}, 198 (2007)
  [arXiv:hep-ph/0610314].

\bibitem{Kouvaris:2006zy}
  C.~Kouvaris, J.~W.~Qiu, W.~Vogelsang and F.~Yuan,
  Phys.\ Rev.\  D {\bf 74}, 114013 (2006)
  [arXiv:hep-ph/0609238].
  
\bibitem{Koike:2007rq} 
  Y.~Koike and K.~Tanaka,
  Phys.\ Rev.\ D {\bf 76}, 011502 (2007)
  [hep-ph/0703169].

\bibitem{Koike:2009ge} 
  Y.~Koike and T.~Tomita,
  Phys.\ Lett.\ B {\bf 675}, 181 (2009)
  [arXiv:0903.1923 [hep-ph]].
  
\bibitem{Zhou:2009jm}
  J.~Zhou, F.~Yuan and Z.~T.~Liang,
  Phys.\ Rev.\  D {\bf 81}, 054008 (2010)
  [arXiv:0909.2238 [hep-ph]].
  
\bibitem{Gamberg:2012iq} 
  L.~Gamberg and Z.~-B.~Kang,
  arXiv:1208.1962 [hep-ph].
  
\bibitem{Metz:2012ui} 
  A.~Metz, D.~Pitonyak, A.~Schaefer, M.~Schlegel, W.~Vogelsang and J.~Zhou,
  arXiv:1209.3138 [hep-ph].
  
\bibitem{Hoyer:2006hu} 
  P.~Hoyer and M.~Jarvinen,
  JHEP {\bf 0702}, 039 (2007)
  [hep-ph/0611293];
  P.~Hoyer, M.~Jarvinen and S.~Kurki,
  JHEP {\bf 0810}, 086 (2008)
  [arXiv:0808.0626 [hep-ph]].

\bibitem{Qian:2011ya} 
  Y.~Qian and I.~Zahed,
  Phys.\ Rev.\ D {\bf 86}, 014033 (2012)
  [Erratum-ibid.\ D {\bf 86}, 059902 (2012)]
  [arXiv:1112.4552 [hep-ph]].
  
\bibitem{Kovchegov:2012ga} 
  Y.~V.~Kovchegov and M.~D.~Sievert,
  Phys.\ Rev.\ D {\bf 86}, 034028 (2012)
  [arXiv:1201.5890 [hep-ph]].
  
\bibitem{Bjorken:1969mm} 
  J.~D.~Bjorken,
  Phys.\ Rev.\ D {\bf 1}, 1376 (1970).
  
\bibitem{Babcock:1978yc} 
  J.~Babcock, E.~Monsay and D.~W.~Sivers,
  Phys.\ Rev.\ Lett.\  {\bf 40}, 1161 (1978);
  Phys.\ Rev.\ D {\bf 19}, 1483 (1979).
  
\bibitem{deFlorian:2009vb} 
  D.~de Florian, R.~Sassot, M.~Stratmann and W.~Vogelsang,
  Phys.\ Rev.\ D {\bf 80}, 034030 (2009)
  [arXiv:0904.3821 [hep-ph]].
    
\bibitem{Ashman:1987hv} 
  J.~Ashman {\it et al.}  [European Muon Collaboration],
  Phys.\ Lett.\ B {\bf 206}, 364 (1988);
  Nucl.\ Phys.\ B {\bf 328}, 1 (1989).
  
\bibitem{Anthony:1996mw} 
  P.~L.~Anthony {\it et al.}  [E142 Collaboration],
  Phys.\ Rev.\ D {\bf 54}, 6620 (1996)
  [hep-ex/9610007].

\bibitem{Abe:1998wq} 
  K.~Abe {\it et al.}  [E143 Collaboration],
  Phys.\ Rev.\ D {\bf 58}, 112003 (1998)
  [hep-ph/9802357].
    
\bibitem{Kuhn:2008sy} 
  S.~E.~Kuhn, J.~-P.~Chen and E.~Leader,
  Prog.\ Part.\ Nucl.\ Phys.\  {\bf 63}, 1 (2009)
  [arXiv:0812.3535 [hep-ph]].
    
\bibitem{Burkardt:2008jw} 
  M.~Burkardt, C.~A.~Miller and W.~D.~Nowak,
  Rept.\ Prog.\ Phys.\  {\bf 73}, 016201 (2010)
  [arXiv:0812.2208 [hep-ph]].
  
\bibitem{Aidala:2012mv} 
  C.~A.~Aidala, S.~D.~Bass, D.~Hasch and G.~K.~Mallot,
  arXiv:1209.2803 [hep-ph].

\bibitem{Jaffe:1991kp}
  R.~L.~Jaffe and X.~D.~Ji,
  Phys.\ Rev.\ Lett.\  {\bf 67}, 552 (1991);
  Nucl.\ Phys.\  B {\bf 375}, 527 (1992).

\bibitem{Tangerman:1994bb}
  R.~D.~Tangerman and P.~J.~Mulders,
  hep-ph/9408305.

\bibitem{Koike:2008du}
  Y.~Koike, K.~Tanaka and S.~Yoshida,
  Phys.\ Lett.\  B {\bf 668}, 286 (2008)
  [arXiv:0805.2289 [hep-ph]].

\bibitem{Lu:2011th}
  Z.~Lu and I.~Schmidt,
  Phys.\ Rev.\ D {\bf 84}, 114004 (2011)
  [arXiv:1109.3232 [hep-ph]].

\bibitem{Metz:2010xs}
  A.~Metz and J.~Zhou,
  Phys.\ Lett.\ B {\bf 700}, 11 (2011)
  [arXiv:1006.3097 [hep-ph]].

\bibitem{Kang:2011jw}
  Z.~B.~Kang, A.~Metz, J.~W.~Qiu and J.~Zhou,
  Phys.\ Rev.\ D {\bf 84}, 034046 (2011)
  [arXiv:1106.3514 [hep-ph]].

\bibitem{Liang:2012rb} 
  Z.~-T.~Liang, A.~Metz, D.~Pitonyak, A.~Schafer, Y.~-K.~Song and J.~Zhou,
  Phys.\ Lett.\ B {\bf 712}, 235 (2012)
  [arXiv:1203.3956 [hep-ph]].
  
\bibitem{Ralston:1979ys} 
  J.~P.~Ralston and D.~E.~Soper,
  Nucl.\ Phys.\ B {\bf 152}, 109 (1979).

\bibitem{Cortes:1991ja} 
  J.~L.~Cortes, B.~Pire and J.~P.~Ralston,
  Z.\ Phys.\ C {\bf 55}, 409 (1992).
    
\bibitem{Kanazawa:2000hz} 
  Y.~Kanazawa and Y.~Koike,
  Phys.\ Lett.\ B {\bf 478}, 121 (2000)
  [hep-ph/0001021];
  Phys.\ Lett.\ B {\bf 490}, 99 (2000)
  [hep-ph/0007272].
    
\bibitem{Kang:2010zzb} 
  Z.~-B.~Kang, F.~Yuan and J.~Zhou,
  Phys.\ Lett.\ B {\bf 691}, 243 (2010)
  [arXiv:1002.0399 [hep-ph]];
  F.~Yuan and J.~Zhou,
  Phys.\ Rev.\ Lett.\  {\bf 103}, 052001 (2009)
  [arXiv:0903.4680 [hep-ph]].
  
\bibitem{Kang:2011ni} 
  Z.~-B.~Kang and F.~Yuan,
  Phys.\ Rev.\ D {\bf 84}, 034019 (2011)
  [arXiv:1106.1375 [hep-ph]].
  
\bibitem{Anselmino:2012rq} 
  M.~Anselmino, M.~Boglione, U.~D'Alesio, E.~Leader, S.~Melis, F.~Murgia and A.~Prokudin,
  arXiv:1207.6529 [hep-ph].

\bibitem{Meissner:2006}
S.~Meissner, Ph.D. thesis, University of Bochum, 2006.

\bibitem{Boer:1997bw} 
  D.~Boer, P.~J.~Mulders and O.~V.~Teryaev,
  Phys.\ Rev.\ D {\bf 57}, 3057 (1998)
  [hep-ph/9710223].
      
\bibitem{Kotzinian:2006dw} 
  A.~Kotzinian, B.~Parsamyan and A.~Prokudin,
  Phys.\ Rev.\ D {\bf 73}, 114017 (2006)
  [hep-ph/0603194].
  
\bibitem{Avakian:2007mv} 
  H.~Avakian, A.~V.~Efremov, K.~Goeke, A.~Metz, P.~Schweitzer and T.~Teckentrup,
  Phys.\ Rev.\ D {\bf 77}, 014023 (2008)
  [arXiv:0709.3253 [hep-ph]].
  
\bibitem{Metz:2008ib} 
  A.~Metz, P.~Schweitzer and T.~Teckentrup,
  Phys.\ Lett.\ B {\bf 680}, 141 (2009)
  [arXiv:0810.5212 [hep-ph]].
  
\bibitem{Wandzura:1977qf} 
  S.~Wandzura and F.~Wilczek,
  Phys.\ Lett.\ B {\bf 72}, 195 (1977).
  
\bibitem{Accardi:2009au} 
  A.~Accardi, A.~Bacchetta, W.~Melnitchouk and M.~Schlegel,
  JHEP {\bf 0911}, 093 (2009)
  [arXiv:0907.2942 [hep-ph]].
  
\bibitem{Kang:2008ey} 
  Z.~-B.~Kang and J.~-W.~Qiu,
  Phys.\ Rev.\ D {\bf 79}, 016003 (2009)
  [arXiv:0811.3101 [hep-ph]];
  Phys.\ Lett.\ B {\bf 713}, 273 (2012)
  [arXiv:1205.1019 [hep-ph]].

\bibitem{Braun:2011aw} 
  V.~M.~Braun, T.~Lautenschlager, A.~N.~Manashov and B.~Pirnay,
  Phys.\ Rev.\ D {\bf 83}, 094023 (2011)
  [arXiv:1103.1269 [hep-ph]].
  
\bibitem{Adler:2004ps} 
  S.~S.~Adler {\it et al.}  [PHENIX Collaboration],
  Phys.\ Rev.\ Lett.\  {\bf 93}, 202002 (2004)
  [hep-ex/0404027];
  Phys.\ Rev.\ D {\bf 73}, 091102 (2006)
  [hep-ex/0602004].
 
\bibitem{Adare:2007dg} 
  A.~Adare {\it et al.}  [PHENIX Collaboration],
  Phys.\ Rev.\ D {\bf 76}, 051106 (2007)
  [arXiv:0704.3599 [hep-ex]];
  Phys.\ Rev.\ Lett.\  {\bf 103}, 012003 (2009)
  [arXiv:0810.0694 [hep-ex]];
  Phys.\ Rev.\ D {\bf 79}, 012003 (2009)
  [arXiv:0810.0701 [hep-ex]];
  Phys.\ Rev.\ D {\bf 83}, 032001 (2011)
  [arXiv:1009.6224 [hep-ex]];
  Phys.\ Rev.\ D {\bf 84}, 012006 (2011)
  [arXiv:1009.4921 [hep-ex]];
  arXiv:1202.4020 [hep-ex].
  
\bibitem{Abelev:2006uq} 
  B.~I.~Abelev {\it et al.}  [STAR Collaboration],
  Phys.\ Rev.\ Lett.\  {\bf 97}, 252001 (2006)
  [hep-ex/0608030];
  Phys.\ Rev.\ Lett.\  {\bf 100}, 232003 (2008)
  [arXiv:0710.2048 [hep-ex]];
  Phys.\ Rev.\ D {\bf 80}, 111108 (2009)
  [arXiv:0911.2773 [hep-ex]].
  
\bibitem{Adamczyk:2012qj} 
  L.~Adamczyk {\it et al.}  [STAR Collaboration],
  Phys.\ Rev.\ D {\bf 86}, 032006 (2012)
  [arXiv:1205.2735 [nucl-ex]].
    
\bibitem{Ageev:2005pq} 
  E.~S.~Ageev {\it et al.}  [COMPASS Collaboration],
  Phys.\ Lett.\ B {\bf 633}, 25 (2006)
  [hep-ex/0511028].
  
\bibitem{Alekseev:2008cz} 
  M.~Alekseev {\it et al.}  [COMPASS Collaboration],
  arXiv:0802.3023 [hep-ex];
  Phys.\ Lett.\ B {\bf 676}, 31 (2009)
  [arXiv:0904.3209 [hep-ex]].
  
\bibitem{Adolph:2012vj} 
  C.~Adolph {\it et al.}  [COMPASS Collaboration],
  arXiv:1202.4064 [hep-ex].
  
\bibitem{Airapetian:1999ib} 
  A.~Airapetian {\it et al.}  [HERMES Collaboration],
  Phys.\ Rev.\ Lett.\  {\bf 84}, 2584 (2000)
  [hep-ex/9907020];
  JHEP {\bf 1008}, 130 (2010)
  [arXiv:1002.3921 [hep-ex]].
  
\bibitem{Adeva:2004dh} 
  B.~Adeva {\it et al.}  [Spin Muon (SMC) Collaboration],
  Phys.\ Rev.\ D {\bf 70}, 012002 (2004)
  [hep-ex/0402010].

\bibitem{deFlorian:2011ia} 
  D.~de Florian, R.~Sassot, M.~Stratmann and W.~Vogelsang,
  Prog.\ Part.\ Nucl.\ Phys.\  {\bf 67}, 251 (2012)
  [arXiv:1112.0904 [hep-ph]].
  
\bibitem{Zhou:2008mz} 
  J.~Zhou, F.~Yuan and Z.~-T.~Liang,
  Phys.\ Rev.\ D {\bf 79}, 114022 (2009)
  [arXiv:0812.4484 [hep-ph]];
  A.~Schafer and J.~Zhou,
  Phys.\ Rev.\ D {\bf 85}, 117501 (2012)
  [arXiv:1203.5293 [hep-ph]].
  
\bibitem{Vogelsang:2009pj} 
  W.~Vogelsang and F.~Yuan,
  Phys.\ Rev.\ D {\bf 79}, 094010 (2009)
  [arXiv:0904.0410 [hep-ph]].
  
\bibitem{Braun:2009mi} 
  V.~M.~Braun, A.~N.~Manashov and B.~Pirnay,
  Phys.\ Rev.\ D {\bf 80}, 114002 (2009)
  [arXiv:0909.3410 [hep-ph]].
    
\bibitem{Ma:2012xn} 
  J.~P.~Ma and Q.~Wang,
  Phys.\ Lett.\ B {\bf 715}, 157 (2012)
  [arXiv:1205.0611 [hep-ph]].
  
\bibitem{Kang:2011hk} 
  Z.~-B.~Kang, J.~-W.~Qiu, W.~Vogelsang and F.~Yuan,
  Phys.\ Rev.\ D {\bf 83}, 094001 (2011)
  [arXiv:1103.1591 [hep-ph]].

\end{thebibliography}
\end{document}